\documentclass[twocolumn,showpacs,amsmath,amssymb,aps,prd,floats,nofootinbib]{revtex4-2}
\usepackage{graphicx}
\usepackage[switch,pagewise,running]{lineno}
\usepackage{dcolumn}
\usepackage{bm}
\usepackage{epsfig}
\usepackage{xspace}
\usepackage{hyperref}
\hypersetup{colorlinks,linkcolor=blue,citecolor=blue,urlcolor=blue}  
\usepackage{color,xcolor}
\usepackage{verbatim}

\begin{document}
\title{Expected Signature For the Lorentz Invariance Violation Effects on $\gamma-\gamma$ Absorption}
\thanks{Corresponding author: Y.G. Zheng, \\ ynzyg@ynu.edu.cn}
\author{Y.G. Zheng$^{1}$, S.J. Kang$^{2}$, K.R. Zhu$^3$, C.Y. Yang$^{4, 5}$, and J.M. Bai$^{4}$}

\affiliation{$^1$Department of Physics, Yunnan Normal University, Kunming, Yunnan, 650092, China; ynzyg@ynu.edu.cn\\
$^2$School of Physics and Electrical Engineering, Liupanshui Normal University, Liupanshui, Guizhou, 553004, China;  kangshiju@alumni.hust.edu.cn\\
$^3$School of Physics and Astronomy, Yunnan University, Kunming, Yunnan, 650092, China\\
$^4$Yunnan Observatories, Chinese Academy of Sciences, Kunming 650011, China\\
$^5$Key Laboratory of Astroparticle Physics of Yunnan Province, Kunming 650091, China}
\begin{abstract}
There are still some {significant and} unanswered questions about the  {incredible}  {very high energy (VHE)} $\gamma$-ray signatures. To help understand the mechanism, focusing on the linear and quadratic perturbation mode for the subluminal regime, the present paper revisited the expected signature for the Lorentz invariance violation effects on $\gamma-\gamma$ absorption in TeV spectra of Gamma-ray bursts (GRBs). We note that there is  {a critical energy} for the pair production process, which is sensitive to the assumed quantum gravity energy scale. We suggest that a {reemergence of the energy spectrum of $\gamma$-rays} at the several tens of TeV is a rough observational diagnostic for the Lorentz invariance violation (LIV) effects. The expected spectra characteristics are applied to a GRB 221009A. The results show that the cosmic opacity with LIV effects considered here  {can} roughly reproduce the observed $\gamma$-ray spectra for the source, which enabled us to constrain the upper limit of the values of energy scale at $E_{\rm QG,~1}\leq3.35\times10^{20}$ GeV for the linear perturbation and $E_{\rm QG,~2}\leq9.19\times10^{12}$ GeV for the quadratic perturbation.  {These scenarios would update the bound of the LIV coefficient with $\xi_{\rm 1}^{\prime}\geq 3.62\times10^{-2}$ for the linear perturbation, and $\xi_{\rm 2}^{\prime}\geq 1.33\times10^{6}$ for the quadratic perturbation in the standard model extension (SME) framework, respectively. }
\end{abstract}

\maketitle

\section{Introduction} \label{sec:intro}
Both the special relativity and the standard model of particle physics are thought to be the low-energy limits of a more fundamental physical theory. In these  {paradigms}, the Lorentz invariance is considered to be a fundamental symmetry in quantum field theory.  {However, efforts to build a fundamental theory, such as quantum gravity or string theories (e.g., \citep[]{1989PhRvD..39..683K,1991NuPhB.359..545K}), suggest a scenario in which the Lorentz invariance is violated at an energy scale thought to be around the \emph{Planck} energy scale with $E_{\rm Planck}\simeq 1.22\times10^{19} \rm GeV$ (e.g., \citep[]{2005LRR.....8....5M,2006AnPhy.321..150J,2010IJMPA..25.5409M,2013LRR....16....5A,2013CQGra..30m3001L}).} Such extreme energies are unreachable to current  {\emph{Earth}-based tests}; most approaches to testing the effect have relied on observations of photons  {travelling} over cosmological distances to accumulate the deviations  {from} Lorentz invariance that may be detectable using the Cherenkov Telescope Array (CTA) (e.g., \citep[]{2017EPJWC.13603018L}), though in the photon sector the $\gamma$-rays observations for the objects in the Galaxy issue that the superluminal effect is likely to inexistent \citep{2022PhRvL.128e1102C}.

A generic approach to Lorentz invariance violation (LIV) effects for photons consists in adding an extra term in their energy-momentum dispersion relation (e.g., \citep[]{1998Natur.393..763A,2018ApJ...865..159A,2019ApJ...870...93A,2016A&A...585A..25T}).  {Despite the} non-infinite value of the hypothetical energy scale, $E_{\rm QG}$, the modified energy-momentum dispersion relation can induce non-negligible observed effects.  {On the one hand, it can cause an energy-dependent photon velocity in a vacuum that, in turn, can translate into an energy-dependent time delay in the arrival time of $\gamma$-ray photons travelling over astrophysical distances \citep{1998Natur.393..763A,2013PhRvD..88h5029E}.} On the other hand, an interesting effect is that the deviation can induce variation in the kinematics of scattering and decay processes (e.g., \citep[]{2003PhRvD..67l4011J}).
As in the case of the pairs production reaction of  {very high energy (VHE)} $\gamma$-rays with photons of the extragalactic background light (EBL), a consequence of the LIV effect allows for reactions that can change energy thresholds (e.g., \citep[]{1999ApJ...518L..21K,2000PhLB..493....1P,2014ApJ...787....4K}), resulting in deviations with respect to standard EBL attenuation in the energy spectrum of distant sources (e.g., \citep[][]{2001APh....16...97S,2008PhRvD..78l4010J}).

It is well known that the LIV effects in the pairs production reaction become relevant at $\gamma$-ray photon energy $E_{\gamma}\sim10~\rm TeV$ in a  {subluminal regime} with  {first-order in the LIV perturbation}. In this scenario, deviations in the scattering kinematics can lead to the reduction of cosmic opacity, thus allowing  {ultrahigh} energy ($\rm E>10~TeV$) $\gamma$-ray photons to arrive at the \emph{Earth} without absorption. Since the upcoming CTA \citep{2013APh....43....3A,2013hegr.confE..33D} and the Large High Altitude Air Shower Observatory (LHAASO) \citep{2014NIMPA.742...95C} will greatly improve the sensitivity above 10 TeV, we expect that the modification of the pairs production reaction of  {VHE} $\gamma$-rays with photons of the EBL can be detected effectively by observations of distant sources at  {VHE}. A feasibility study on the possible detection of LIV effects in the observed spectral of sources has been performed on the bias of the expected opacity for different values of the hypothetical energy scale, $E_{\rm QG}$\citep{2014JCAP...06..005F,2016A&A...585A..25T,2018ApJ...865..159A,2019ApJ...870...93A}.

The present paper discusses the reduction of the EBL pairs production opacity due to the LIV effect on cosmological photon propagation along the line of sight to distant sources. We expected to find the signature for the LIV effects on $\gamma-\gamma$ absorption. The paper is structured as follows. In Section 2, we present a detailed description of the cosmic opacity with LIV. In Section 3, we apply the EBL pairs production opacity with the LIV effect to the high energy  {hardening} spectral phenomena of distant sources. In Section 4, we discuss the results. Throughout the paper, we assume the Hubble constant $H_0=67.4\rm{ \; km \; s^{-1} \; Mpc^{-1}}$, the dimensionless cosmological constant $\Omega_{\Lambda}=0.685$, matter energy density $\Omega_{M}=0.315$, and radiation energy density $\Omega_{R}=0$, respectively \citep{2020A&A...641A...6P}.

\section{Lorentz Invariance Violation} \label{sec:dataset}
Quantum gravity models (e.g., \citep[]{1989PhRvD..39..683K,1991NuPhB.359..545K}) which assume LIV imply that there is a modification in  {the energy-momentum dispersion relation \citep{1998Natur.393..763A,2006APh....25..402E}} for a particle of mass,  {$m_{\rm i}$},
\begin{equation}
E_{\rm i}^{2} \simeq p_{\rm i}^{2}c^{2}+m_{\rm i}^{2}c^{4}+{s_{\rm{n}}} E_{\rm i}^{2}(\frac{E_{\rm i}}{E_{QG}})^{n}\,,
\end{equation}  where  {$E_{\rm i}$ and $p_{\rm i}$} are the energy and momentum of the particle, $c$ is the conventional velocity of light, $n$ is the leading order of the LIV perturbation, and  {${s_{\rm{n}}}$} shows that this perturbation is model-dependent and refers to  {superluminal} ( {${s_{\rm{n}}}=+1$}, increasing photon speed with increasing energy) and subluminal ( {${s_{\rm{n}}}=-1$}, decreasing photon speed with increasing energy) scenarios. Considering the sensitivity of current detectors, only the linear, $n=1$, and/or quadratic, $n=2$, perturbation modes  {are} taken into account for constraints on the hypothetical energy scale, $E_{\rm QG}$ (e.g., \citep[]{2016CRPhy..17..632H}).

\section{Cosmic Opacity with LIV}
\begin{figure*}
\centering
\includegraphics[height=12cm,width=18cm]{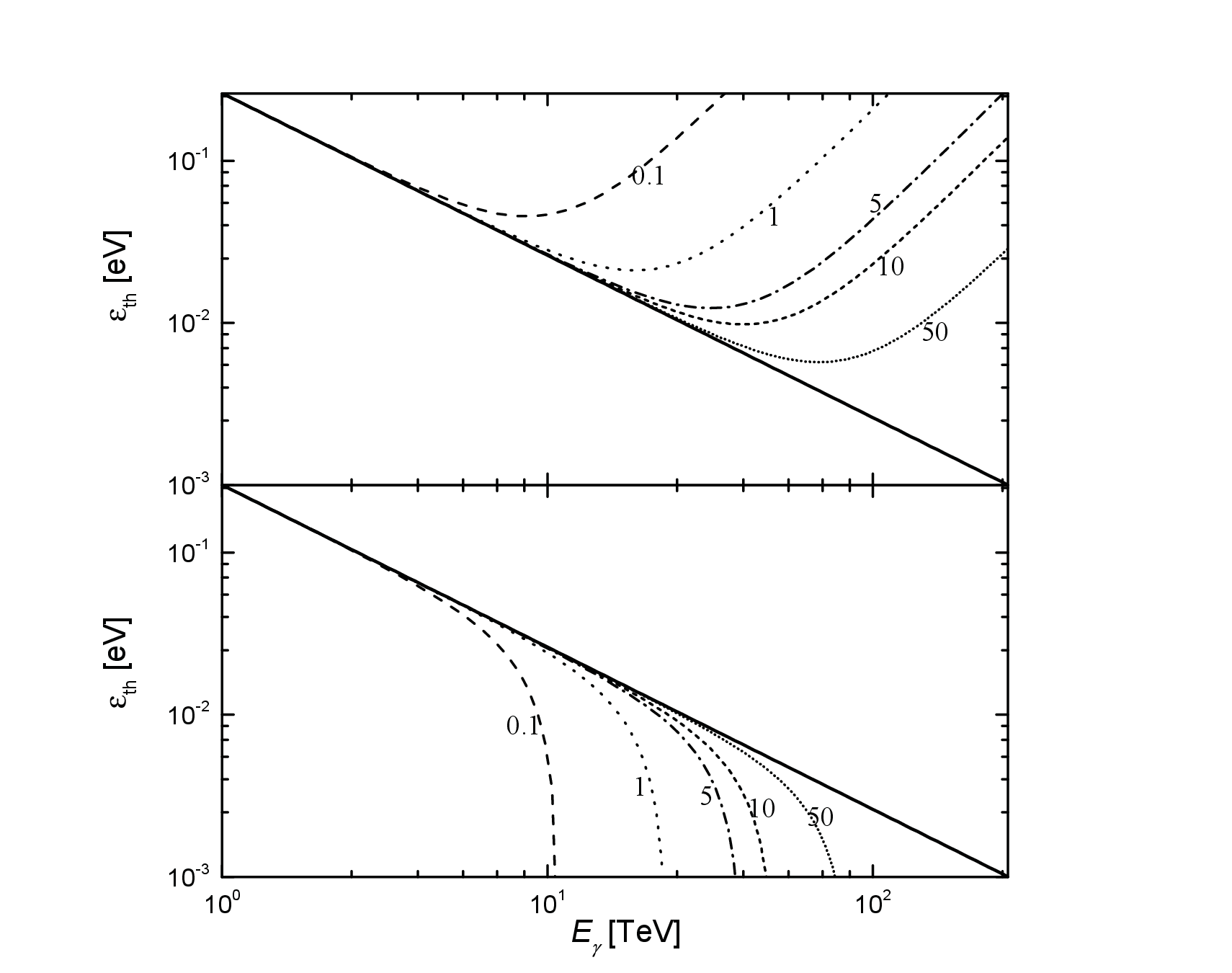}
\caption{Threshold energy of the target photon for the pair production as a function of the $\gamma$-ray photon energy in the regime of the linear perturbation. The black solid line shows the standard case. The other lines indicate the modified threshold, which results from the LIV-modified kinematics, for different values of the quantum gravity energy scale. Marks near the curves represent the quantum gravity energy scale in units of Plank energy. The top panel shows the subluminal regime, and the bottom panel shows the superluminal regime.}
\label{sub:fig1}
\end{figure*}

Considering the LIV effect in the photon sector, the specific energy-momentum dispersion relations can be written as
\begin{equation}
E_{\gamma}^{2} \simeq p_{\gamma}^{2}c^{2}+{s_{\rm{n}}} E_{\gamma}^{2}(\frac{E_{\gamma}}{E_{QG}})^{n}\,.
\end{equation}
The threshold energy of the pair production process is determined by the relation \citep{1999ApJ...518L..21K}
\begin{equation} 
E_{\rm total}^{2}-p_{\rm total}^{2}c^{2}\geq(2m_{e}c^{2})^{2}\,,
\end{equation}
where $E_{\rm total}$ and $p_{\rm total}$ total energy and momentum of the initial state, $m_{e}$ is the electron mass. Assuming two photons be from opposite directions with collinear final momenta, we can deduce  the pair production threshold energy, $\epsilon_{\rm th}$, that is the threshold of the target photon (e.g., \citep[]{2016A&A...585A..25T,2018ApJ...865..159A,2019ApJ...870...93A}),
\begin{equation}
\epsilon_{\rm th}=\frac{m_{e}^{2}c^{4}}{E_{\gamma}}-{s_{\rm{n}}} \frac{E_{\gamma}^{(n+1)}}{4E_{QG}^{n}}\,,
\end{equation}  
where the second term in the right is introduced by LIV effects. Figure \ref{sub:fig1} plots the target photon energy threshold for the pair production as a function of the $\gamma$-ray photon energy in the regime of the linear perturbation. The black solid line shows the standard case $\epsilon_{\rm th}=m_{e}^{2}c^{4}/E_{\gamma}$. The other lines indicate the modified threshold, which results from the LIV-modified kinematics, for different values of the quantum gravity energy scale, $E_{\rm QG}$.

\begin{figure*}
\centering
\includegraphics[height=7cm,width=13cm]{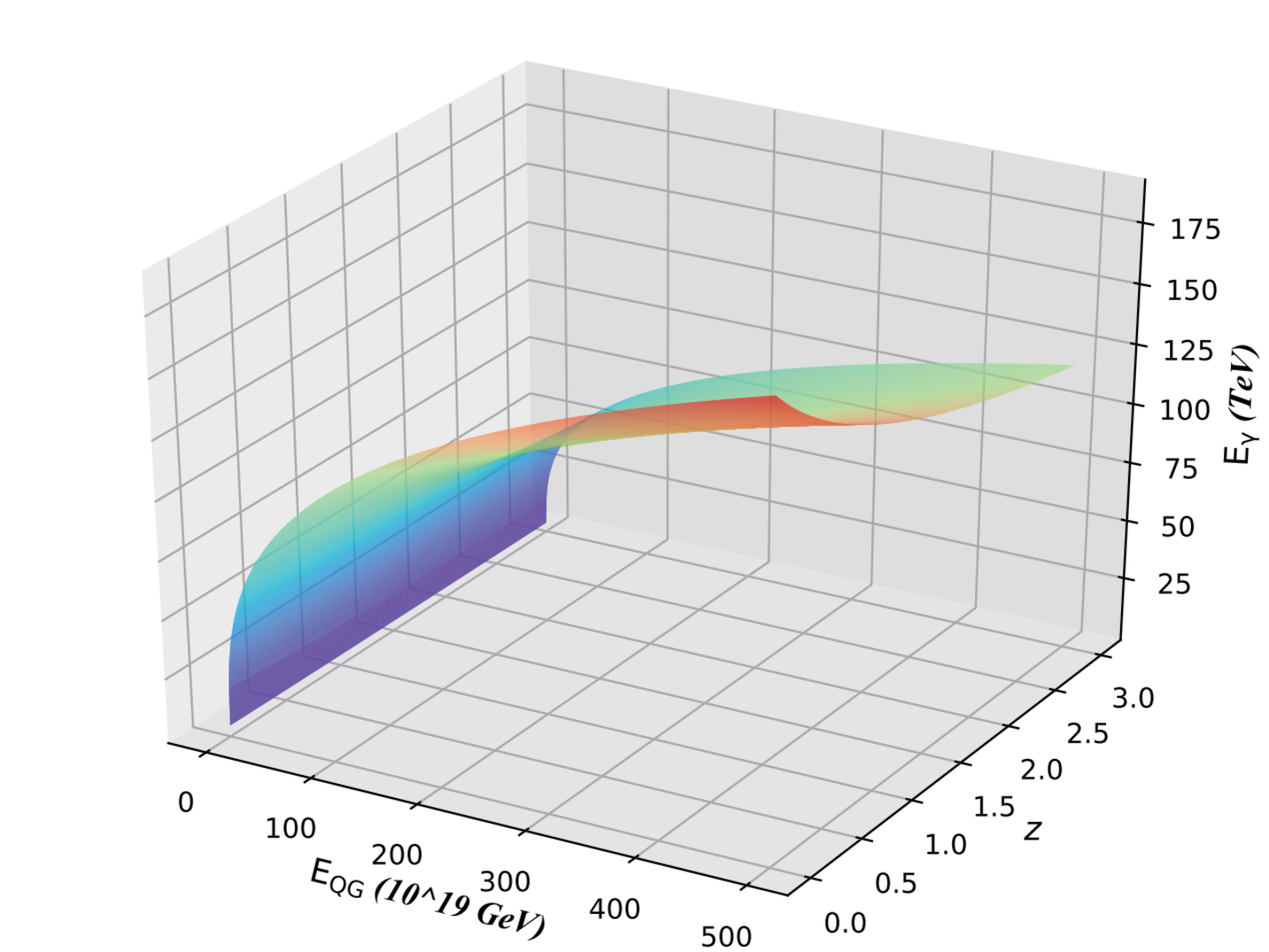}
\caption{ {The critical energy} for the pair production as a function of the quantum gravity energy scale in the linear perturbation mode for the subluminal regime in the observed frame.}
\label{sub:fig2}
\end{figure*}

As shown in Figure \ref{sub:fig1}, an attractive feature of the curves for the subliminal regime is the existence of a {minimum photon energy threshold}.  {It implies a progressive reduction of the resulting optical depth above critical energy, $E_{\gamma,\rm cr}=\left[-4m_{e}^{2}c^{4}E_{\rm QG}^{n}/s_{\rm n}(n+1)\right]^{1/(n+2)}$.}
That is, $\gamma$-rays of these energies are non-effectively absorbed through interaction with the low energy radiation of the EBL. 
We show {the critical energy} for the pair production as a function of the quantum gravity energy scale in the linear perturbation mode for the subluminal
regime in the observed frame in Figure \ref{sub:fig2}. It can be seen that  {the critical energy} for the pair production is sensitive to the quantum gravity energy scale.
Since the $\gamma$-ray decay process  dominated  the superluminal regime (e.g., \citep[]{2021PhRvD.104f3012L}), we concentrate on the following treatment to the subluminal regime with $s_{\rm n}=-1$.

 Considering the framework of the standard model extension (SME) \cite{1997PhRvD..55.6760C,1998PhRvD..58k6002C} with dimension-5 charge parity-time reversal (CPT) operators \citep{2003PhRvL..90u1601M}, an effective field theoretic (EFT) treatment of LIV, where the dispersion relation of the photon field is nothing but an analogy to the generalized relation (1) in the context by the replacement $\frac{s_{\rm n}}{E^n_{\rm QG}}\rightarrow \pm \frac{\xi_{\rm n}}{E^n_{\rm planck}}$.
 In this scenario, the Eq. (2) and (4) {can be rewritten} as $E_{\gamma}^{2} \simeq p_{\gamma}^{2}c^{2}\pm\xi_{\rm n} E_{\gamma}^{2}(\frac{E_{\gamma}}{E_{\rm planck}})^{n}$, and $\epsilon_{\rm th} = \frac{m_{e}^{2}c^{4}}{E_{\gamma}}\mp\xi_{\rm n} \frac{E_{\gamma}^{(n+1)}}{4E_{\rm planck}^{n}}$, respectively. The observed VHE spectra can be used to evaluate the related limits to the LIV coefficient $\xi_{\rm n}^{\prime}$ entailed in the SME framework with a parameterized relationship $\xi_{\rm n}^{\prime}=E_{\rm planck}/E_{\rm QG}=\xi_{\rm n}^{1/n}$ (e.g., \citep[]{2018ApJ...865..159A}), where the dimensionless parameter $\xi_{\rm n}^{\prime}$ depends on particle type and theoretical framework (e.g., \citep[]{1998Natur.393..763A,2016A&A...585A..25T}).

We assume that LIV effects {do} not change the functional form of the pair production cross-section. In this scenario, the optical depth at the energy $E_{\gamma}$ and for a TeV photon from a source at redshift $z$ can be evaluated as
\begin{eqnarray}
\label{eq5}
{\tau _\gamma }({E_\gamma },\;z) = c\int_0^z {(\frac{{{\text{d}}t}}{{{\text{d}}z'}})} {\text{d}}z'\int_{ - 1}^{ + 1} {\text{d}} \mu \;\frac{{1 - \mu }}{2} \nonumber \\
\times \int_{\varepsilon _{{\text{th}}}^\prime }^\infty  {d{\varepsilon ^\prime }{n_\varepsilon }({\varepsilon ^\prime },{z^\prime })} {\sigma _{\gamma \gamma }}(E_\gamma ^\prime ,{\varepsilon ^\prime },\mu )
\end{eqnarray}
where $\mu\equiv cos\theta$ with the angle, $\theta$, between the EBL photon of energy $\epsilon$ and $\gamma$-ray photon, $n_{\epsilon}(\epsilon',\ z')d\epsilon'$ is the comoving number density of EBL photons with energies between $\epsilon'$ and $\epsilon'\,+\,d\,\epsilon'$ at redshift $z'$, $\epsilon'_{\rm th}=\epsilon_{\rm th}(E'_{\gamma},\ \mu)$, $E'_{\gamma}= E_{\gamma}(1+z')$, and $\sigma_{\gamma \gamma}$ is the total pair production cross section. For a flat universe, the differential of time to be used in Eq. \ref{eq5} is
\begin{equation}
\label{eq6}
\frac{d t}{d z'}=\frac{1}{(1+z')H_{0}}[(1+z')^{2}(1+\Omega_{M}z')-z'(z'+2)\Omega_{\Lambda}]^{-1/2}\,.
\end{equation}

\begin{figure*}
\centering
\includegraphics[height=7cm,width=13cm]{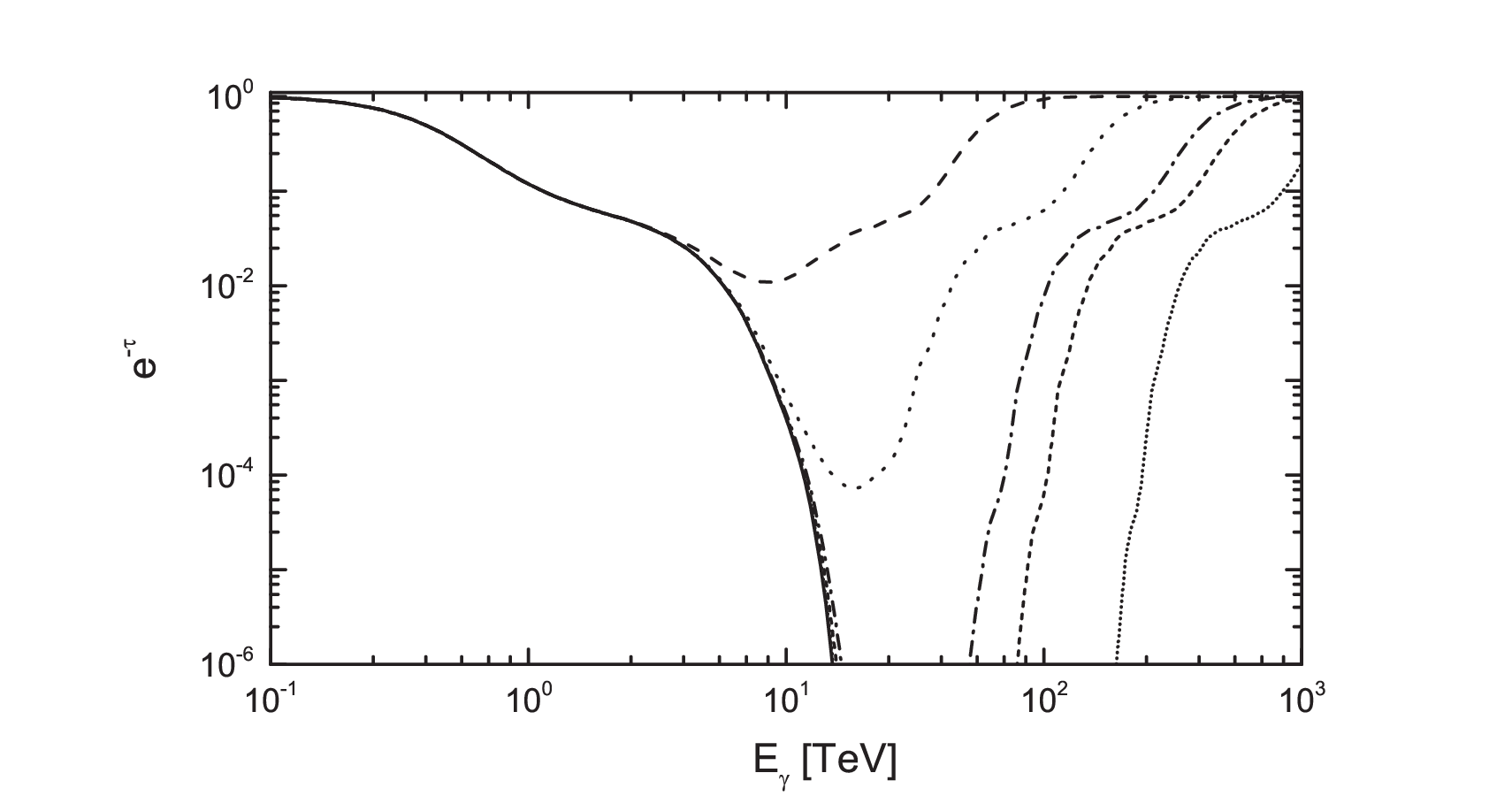}
\caption{Absorption coefficient $e^{-\tau}$ as a function of energy for the $\gamma$-rays that propagate from a source at redshift $z=0.02$, using the EBL photon densities of Ref. \cite{2017A&A...603A..34F,2018A&A...614C...1F}. The black solid line indicates the case of standard QED, the dashed line shows the modified coefficient for $E_{\rm QG}=0.1E_{\rm Planck}$, the dotted line shows the modified coefficient for $E_{\rm QG}=E_{\rm Planck}$, the dash-dotted line shows the modified coefficient for $E_{\rm QG}=5E_{\rm Planck}$, the short dashed line shows the modified coefficient for $E_{\rm QG}=10E_{\rm Planck}$, and the short dotted line shows the modified coefficient for $E_{\rm QG}=50E_{\rm Planck}$.}
\label{sub:fig3}
\end{figure*}

Using the EBL photon densities of Ref. \cite{2017A&A...603A..34F,2018A&A...614C...1F}, we calculate the optical depth for $\gamma$-ray photons from a source at redshift $z=0.02$.  {The comparison with the absorption coefficient of the standard case and the absorption coefficient of the LIV effect is presented in Figure \ref{sub:fig3}.} One can see that the drastic reduction in the opacity above a few tens of TeV induced by the LIV effect is clearly visible.  {Since the number density of the low-energy EBL photons decreases with energy, there are less background lights to interact with the energetic photon above the critical energy $E_{\gamma,\rm cr}$ and, therefore, a {reemergence of the energy spectrum of  $\gamma$-rays} \citep{2010MPLA...25.3251S} can be expected.} It should be emphasized {a reemergence of the energy spectrum of  $\gamma$-rays} in the several tens of TeV is significant imprint of LIV effects. 
It can provide a rough energy spectrum observational diagnostic for the LIV effects.

\section{Application to VHE Spectra Characteristic} 
Due to the bright $\gamma$-ray luminosity above several tens of TeV, at which the LIV effects become fully appreciable, the gamma-ray bursts (GRBs) can be treated as an ideal source to test possible {a reemergence of the energy spectrum of  $\gamma$-rays} induced by LIV effects. It commonly exhibits typical characteristics that the spectra  {are softening} with energy increasing as a result of decreasing inverse Compton cooling and/or particle acceleration efficiencies. This display is in conflict with the requirement. We  {can} no longer ignore the fact that the energy of the highest photon of GRB 221009A is detected up to 18 TeV by LHAASO (e.g., \citep{Huang32677}) and  up to 251 TeV by Carpet 2 (e.g., \citep{2022ATel15669....1D,2022ATel15675....1F}), which are particularly suitable for the present analysis (e.g., \citep{2023ApJ...942L..21F}).

\begin{figure*}[htbp]
\centering
 \includegraphics[height=6cm,width=8.5cm]{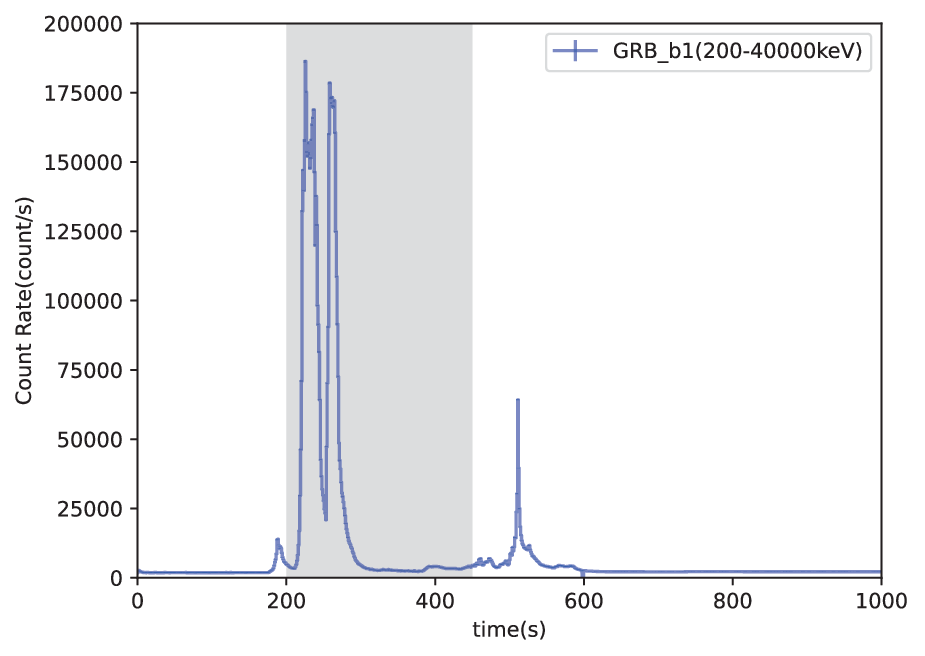}
  \includegraphics[height=6cm,width=8cm]{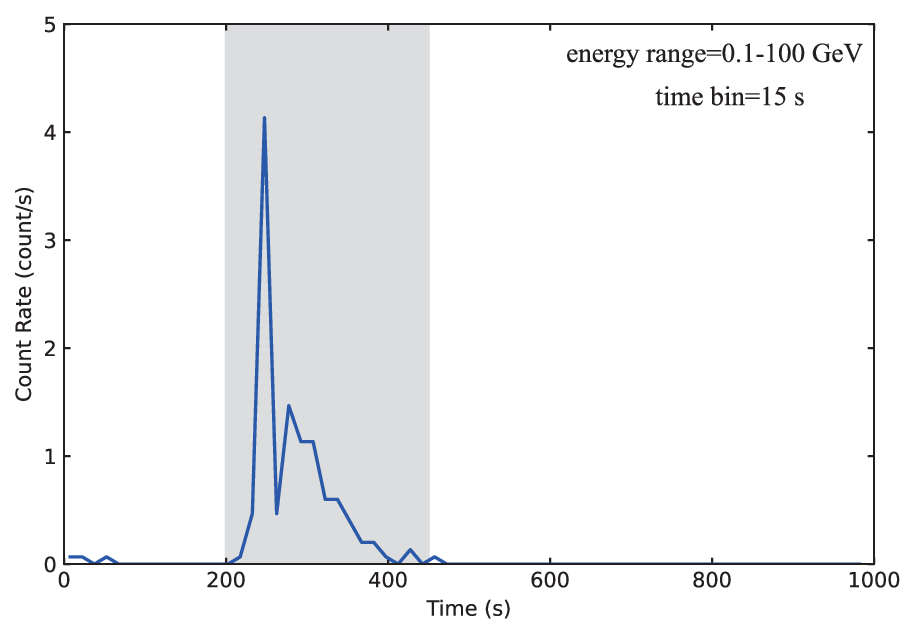}
 \caption{The Light Curve of GRB 221009A  at 0-1000s after burst observed by Fermi-GBM (left)and Fermi-LAT (right) respectively. The left panel corresponds to energy range 200-40000 keV,
while right panel represents the 15 seconds binned light curve in the 0.1-100 GeV energy range.
 }
 \label{fig:LC}
\end{figure*}

\begin{figure*}
\centering 
\includegraphics[width=14cm,height=10cm]{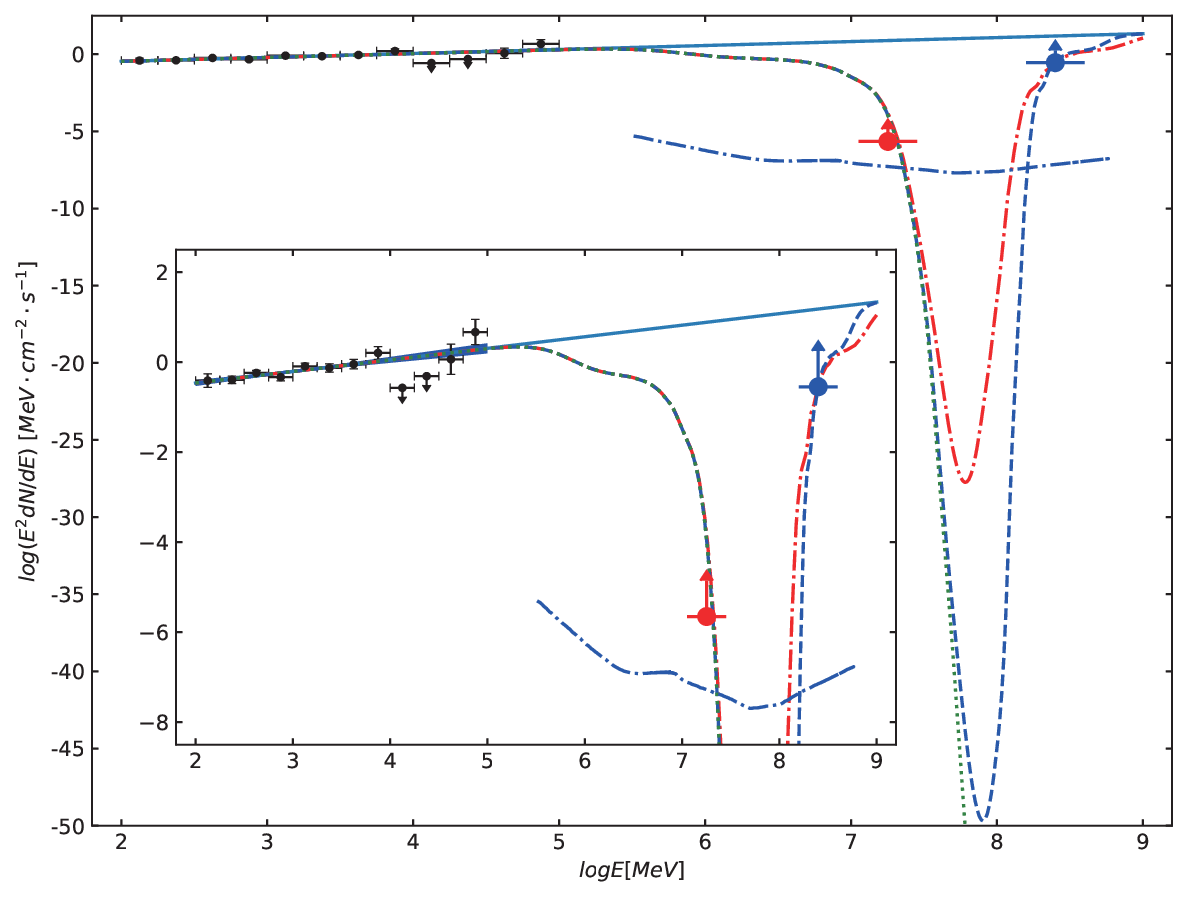}
\caption{The spectral energy distributions (SEDs) of GRB 221009A. The black solid circles represent the Fermi-LAT spectrum obtained from 200s - 450s after the burst. The red dot indicates the observed lower limit flux of LHAASO at 18 TeV. The blue dot represents 251 TeV photons 95\% lower limit flux estimated from this burst by the Carpet 2 detector after the GBM trigger \cite{2023ApJ...942L..21F}. The solid lines indicate the intrinsic spectrum of the burst that extrapolates the obtained 0.1-100 GeV LAT spectrum. The EBL attenuation spectrum and the model spectra in the case of cosmic opacity with LIV effects for both the linear and quadratic perturbations are indicated by green dotted, red dot-dashed, and blue dashed lines, respectively. The blue dashed-dotted line represents the differential sensitivity curve of LHAASO for one year of exposure time \cite{2019arXiv190502773C}.}
\label{fig:SED}
\end{figure*}

GRB 221009A is an extremely bright and long-duration GRB with a redshift of z=0.151 (e.g., \citep[]{2022GCN.32688....1K,2022GCN.32928....1I})  or z = 0.15 (e.g., \citep[]{2022ATel15656....1P,2022GCN.32800....1D,2022GCN.32648....1D}) and entered at RA = 288.264, Dec = 19.773. It was detected by the Swift Burst Alert Telescope (BAT) instruments (e.g., \citep[]{2022GCN.32632....1D,2022GCN.32635....1K}) at T0 = 2022-10-09 13:16:59 (UTC), the Fermi Gamma-Ray Burst Monitor (GBM) instruments \cite{2022GCN.32636....1V}, and the Fermi Large Area Telescope (Fermi-LAT) instruments \cite{2022GCN.32637....1B} on 2022 October 9. The Fermi-GBM light curve are shown in Figure \ref{fig:LC}, left panel. From the LAT Data Serve, we selected the higher energy events of GRB 221009A at 0-1000s after the burst. The events count curve of GRB 221009A in $3$ ROI is shown in Figure \ref{fig:LC}, right panel. Selecting events during the GRB 221009A active period, where it is located at 200s - 450s after burst and shows as a grey area in Figure 4, we used the Fermipy platform \citep{2017ICRC...35..824W} for Fermi spectral analysis. Since the highest energy of the event is up to about 99.3 GeV, the energy range of Fermi Likelihood analysis is adopted from 0.1 to 100 GeV. The result of the Fermi Likelihood analysis shows that the spectrum is described as a power-law, given by
\begin{equation}
\frac{dN}{dE} = N_{0}(\frac{E}{E_{0}})^{\Gamma}
\label{Eq:8}
\end{equation}
with the characteristic energy $E_{0} =1 \rm GeV$, the spectral index $\Gamma= -1.74\pm 0.06$ and the prefactor $N_{0}=(6.22\pm0.46)\times 10^{-7}~\rm MeV^{-1}~cm^{-2}~s^{-1}$, and the test statistic (TS) value is 1264.84. Dividing the observed energy band into 12 energy bins on average, we obtain  {the spectral energy} distributions (SEDs) from 0.1 to 100 GeV.

In Figure \ref{fig:SED}, we plot the SED from 0.1 to 100 GeV and Fermi best-fitted lines of GRB 221009A. Figure \ref{fig:SED} also compiles the observed lower limit flux of LHAASO at 18 TeV and Carpet 2 at 251 TeV from Ref. \cite{2023ApJ...942L..21F}. From the phenomenological and theoretical point of view, the source GRB 221009A should make good targets for the effective investigation of LIV effects spectral anomalies despite the fact that the peculiar emission properties of  {these sources are still not clearly understood}. It is believed that the primary TeV photons propagating through intergalactic space should be attenuated due to their interactions with the EBL to produce electron-positron ($e^{\pm}$) pairs (e.g., \citep[]{1966PhRvL..16..252G,1992ApJ...390L..49S,2012Sci...338.1190A,2013APh....43..112D,2013A&A...554A..75S}). The relatively large redshift of GRB 221009A implies an important absorption of the VHE spectrum.

Application of the model to study whether the LIV effects spectral anomalies can give an excellent fitting for the VHE $\gamma$-ray photon spectrum, we extrapolate the obtained 0.1-100 GeV LAT spectrum from this burst to higher energies and use this to limit the intrinsic spectrum of the burst.  {Considering} the absorption effect of EBL, the photon spectrum observed at the Earth becomes
\begin{equation}
\frac{dN_{\rm \gamma}^{\rm obs}}{dE_{\rm \gamma}}=\frac{dN_{\rm \gamma}^{\rm int}}{dE_{\rm \gamma}}\exp[-\tau_{\rm \gamma}(E_{\rm \gamma},z)]\;.
\label{eq:8}
\end{equation}
In Figure {\ref{fig:SED}}, we also include the EBL attenuation spectrum (green dotted lines) and the model spectra in the case of cosmic opacity with LIV effects in both the linear (red dotted-dashed lines) and quadratic (blue dashed lines) perturbation mode for the subluminal regime. It is interesting to compare the model spectra for the case of standard EBL attenuation with the case of cosmic opacity with LIV effects. In the case of standard EBL attenuation, the theoretical $\gamma$-ray spectra display a power-law shape at low energies, followed by an exponential decrease at TeV energies. Alternatively, in the regime of cosmic opacity with LIV effects, a convex spectrum can be seen around  {the critical energy}, $E_{\gamma,\rm cr}$. We note that  {a reemergence of the energy spectrum of  $\gamma$-rays} for the case of cosmic opacity with LIV effects in consequence of the drastic reduction in the cosmic opacity, leading up to the model spectra  {transforming} smoothly from the case of standard EBL attenuation to the intrinsic spectrum.

It is clear from Figure {\ref{fig:SED}} that the cosmic opacity with LIV effects considered here is able to roughly reproduce the observed $\gamma$-ray spectra for the GRB 221009A. The predicted spectra for the EBL model give the energy scale $E_{\rm QG,1}\leq3.35\times10^{20}$ GeV for the linear perturbation and $E_{\rm QG,2}\leq9.19\times10^{12}$ GeV for the quadratic perturbation. These values corresponds to a critical energy $E_{\rm \gamma,~cr,~1}\simeq 55.95~\rm TeV$ for the linear and $E_{\rm \gamma,~cr,~2}\simeq 73.66~\rm TeV$ for the quadratic in the observed frame, respectively. The determined energy scale satisfies the condition $\xi_{\rm 1}^{\prime}\geq 3.62\times10^{-2}$ and $\xi_{\rm 2}^{\prime}\geq 1.33\times10^{6}$.

\section{Discussion and Conclusion} \label{sec:Conclusion and Discussion}
As an open issue, the dedicated experimental tests of LIV effects can help to clear the path to a unification theory of the fundamental forces of nature. Since some LIV effects are expected to increase with energy and over long distances due to cumulative processes in photon propagation, astrophysical searches provide sensitive probes of LIV effects and their potential signatures (e.g., \citep[]{2001APh....16...97S,2003APh....20...85S,2008PhRvD..78h5026K,2009PhRvD..80c6010H,2013PhRvD..87l2001V,2017PhRvD..95f3001M,2019EPJC...79.1011S,2019JCAP...04..054A}).
Focusing on the linear perturbation mode for  {the subluminal regime}, the present paper revisited the expected signature for the LIV effects on $\gamma-\gamma$ absorption in TeV spectra of distant sources. We note that the existence of a {minimum photon energy threshold} for the pair production process leads up to a  {critical energy}, which is sensitive to the quantum gravity energy scale. We suggest that  {a reemergence of the energy spectrum of $\gamma$-rays} in the few tens of TeV is a rough observational diagnostic for the LIV effects. The expected spectra characteristics are applied to a surprising source, GRB 221009A.  The results show that the cosmic opacity with LIV effects considered here is able to roughly reproduce the observed $\gamma$-ray spectra for the source.

The early efforts propose that the best constraint from the current data is $\xi_{\rm 1}^{\prime}\leq O(1000)$ (e.g., \citep[]{1999PhRvL..83.2108B,1999PhRvL..82.4964S}), far below the natural order with $\xi_{\rm 1}^{\prime}=1$ as a physically best-motivated choice is expected (e.g., \citep[]{2009ARNPS..59..245L,2014JCAP...06..005F,2016A&A...585A..25T}). The recent, analysing on $\gamma$-ray burst (GRB) photon propagation results in a suggestion of a subluminal light speed variation in a vacuum with the linear perturbation energy scale determined to be $E_{\rm QG,1}\simeq3.6\times10^{17}$ GeV \citep{2016APh....82...72X,2016PhLB..760..602X, 2018JCAP...01..050X,2021PhLB..82036518Z}, corresponding to a large LIV coefficient with $\xi_{\rm 1}^{\prime}\simeq33.9$. It is noted that there is some previous attempt to constrain the bounds of the LIV coefficient $\xi_{\rm n}^{\prime}$. On the upper limited side, reference \cite{2013PhRvD..87l2001V} find $\xi_{\rm 1}^{\prime}< 0.39$ and $\xi_{\rm 2}^{\prime}< 10^9$ from time-of-flight measurements of photons from GRBs. Reference \cite{2019PhRvD..99d3015L} found 2$\sigma$ upper limits $\xi_{\rm 1}^{\prime}< 0.1$ and $\xi_{\rm 2}^{\prime} <5.26\times10^6$ using VHE $\gamma$-ray spectra of blazars detected by imaging atmospheric Cherenkov telescopes. On the other side, based on the modified dispersion relation for photons in the subluminal regime, reference \cite{2023ApJ...942L..21F} impose the pair production process on the observed lower limit of the LHAASO and Carpet 2 derived from GRB 221009A. The deduced lower limit values of the LIV coefficient are the same order of magnitude as our current results, which is significantly less than the early efforts (e.g., \citep[]{2009ARNPS..59..245L,2014JCAP...06..005F,2016A&A...585A..25T,2016APh....82...72X,2016PhLB..760..602X,2018JCAP...01..050X,2021PhLB..82036518Z}). 

A potential approach to test the LIV effects due to cumulative processes in photon propagation is to study whether the photon index above several tnes of TeV significantly  {decreases}. We argue that if the VHE emission from distant sources  {is} dominated by inverse Compton scattering of the relativistic electrons on the soft photon, the LIV effects spectral anomalies should result in spectral hardening at several tens of TeV. Nevertheless, some other possible explanations of the unexpected VHE $\gamma$-ray signatures seen in a few individual distant sources include the hypothesis, such as lower EBL density than expected from the current EBL model (e.g., \citep[]{2013ApJ...768L..31F}) and/or EBL inhomogeneities (e.g., \citep[]{2013ApJ...768L..31F,2017MNRAS.467.2896K}), an additional $\gamma$-ray emission component (e.g., \citep[]{2010APh....33...81E,2013ApJ...764..113Z,2015JPhCS.632a2035D,2016A&A...585A...8Z}), and the  {existence} of exotic anion-like particles into which enable VHE photons to avoid absorption (e.g., \citep[]{2011JCAP...11..020D,2017A&A...603A..59D,2020PhRvD.101f3004L,2021PhRvD.104h3014L}).

It is attempt to constrain on the energy scale of LIV using two independent channels: a temporal approach considers the possibility of energy dependence in the arrival time of $\gamma$-rays, whereas a spectral approach considers the possibility of modifications  {in the} interaction of VHE $\gamma$-rays with EBL photons (e.g., \citep[]{2016A&A...585A..25T,2019ApJ...870...93A}). The present work differs from the earlier studies that treat the energy scale of LIV as a free parameter. We expect a distinctive energy spectra upturn from standard EBL absorption effects.  {It is also worth noticing that the TeV $\gamma$-ray  {a reemergence of the energy spectrum of $\gamma$-rays} could provide a bound constraint for the concentrative subluminal regime in the SME framework. We can predict the critical energy $E_{\rm \gamma, ~cr, ~1}\simeq 18.5(\xi_{\rm 1}^{\prime})^{-1/3}~\rm TeV$ for the linear and $E_{\rm \gamma, ~cr, ~2}\simeq 84.8(\xi_{\rm 2}^{\rm\prime 2})^{-1/4}~\rm PeV$ for the quadratic, respectively. While the critical energy are usually limited by the maximal energy of the experiment measured, in the context, we may have a chance to assume that the critical energy is in the range of $18\rm~TeV \leq E_{\rm \gamma, ~cr}\leq 251\rm~TeV$. The assumptions constraint on the bounds of $\xi_{\rm n}^{\prime}$ turns to be $1.09\geq\xi_{\rm 1}^{\prime}\geq 4.55\times10^{-4}$ for the linear perturbation, and $2.22\times10^{7}\geq\xi_{\rm 2}^{\prime}\geq 1.24\times10^{5}$ for the quadratic perturbation. The fact of that the rarely observed energy spectra of tens of TeV $\gamma$-ray results in an open issue for the critical energy. We could expect the possibility that some of the upcoming TeV-PeV events come from distant sources. We leave a wider discussion of this type of SME bounds for future experiments.
}

\acknowledgements
We thank the anonymous referee for their valuable comments and suggestions. We thank Dr J.M. Chen for the \emph{Fermi} light curve data processing. This work is partially supported by the National Natural Science Foundation of China (Grant Nos.11873043, 12163002, U2031111, and 12233006), and the Research Foundation of Liupanshui Normal University (LPSSYKJTD201901).

\appendix

\bibliographystyle{apsrev}
\bibliography{article.bib}
\end{document}